# An efficient floating point multiplier design for high speed applications using Karatsuba algorithm and Urdhva-Tiryagbhyam algorithm

Arish S
School of VLSI Design and Embedded Systems
National Institute of Technology Kurukshetra
Kurukshetra, India
arishsu@gmail.com

R.K.Sharma
School of VLSI Design and Embedded Systems
National Institute of Technology Kurukshetra
Kurukshetra, India
rksharama@nitkkr.ac.in

**Abstract:** Floating point multiplication is a crucial operation in high power computing applications such as image processing, signal processing etc. And also multiplication is the most time and power consuming operation. This paper proposes an efficient method for IEEE 754 floating point multiplication which gives a better implementation in terms of delay and power. A combination of Karatsuba algorithm and Urdhva-Tiryagbhyam algorithm (Vedic Mathematics) is used to implement unsigned binary multiplier for mantissa multiplication. The multiplier is implemented using Verilog HDL, targeted on Spartan-3E and Virtex-4 FPGA.

Keywords: fpga, Floating point multiplier, Vedic mathematics, Urdhva-Tiryagbhyam, Karatsuba

## I. INTRODUCTION

Floating point multiplication units are an essential IP for modern multimedia and high performance computing such as graphics acceleration, signal processing, image processing etc. There are lot of effort is made over the past few decades to improve performance of floating point computations. Floating point units are not only complex, but also require more area and hence more power consuming as compared to fixed point multipliers. And the complexity of the floating point unit increases as accuracy becomes a major issue. IEEE 754 [1] support different floating point formats such as Single Precision format, Double Precision format, Quadruple Precision format etc. But as the precision increases, multiplier area, delay and power increases drastically. In the proposed paper, we present a new multiplication method which uses a combination of Karatsuba and Urdhva-Tiryagbhyam (Vedic Mathematics) algorithm for multiplication. This combination not only reduces delay, but also reduces the percentage increase in hardware as compared to conventional methods.

IEEE 754 format specifies two different formats namely single precision and double precision format [1, 2]. Fig. 1 shows the different IEEE 754 floating point formats used commonly. The Single precision format is of 32-bit wide and Double precision format is of 64-bit wide. The Most

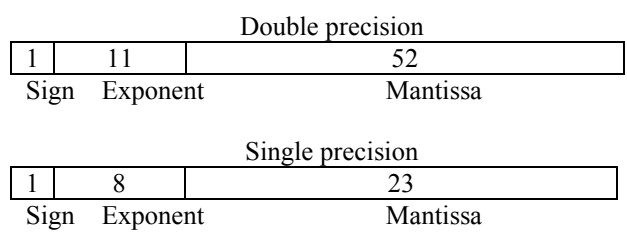

Fig. 1 Floating point formats in the proposed model

Significand Bit is the sign bit. The exponent is a signed integer. It is often represented as an unsigned value by adding a bias. In
Single precision format, the exponent is of 8-bit wide and the bias is 127, i.e. the exponent has a range of $(-127 \text{ to } 128)$. In Double precision format, the exponent is of 11-bit wide and the bias is 1023, i.e. the exponent has a range of $(-1023 \text{ to } 1024)$. The mantissa or significand of Single precision format is of 23-bit and of double precision format is of 52 bit wide. The maximum value that can be represented using floating point format is
$$largest\ significand \times base^{largest\ exponent}.$$
And the minimum value that can be represented is
$$smallest\ significand \times base^{smallest\ exponent}.$$

## II. FLOATING POINT MULTIPLIER DESIGN

A floating point number has four parts: sign, exponent, significand or mantissa and the exponent base. A floating point number is represented in IEEE-754 format [1, 2] as $\pm s \times b^e$ or $\pm significand \times base^{exponent}$. The exponent base for binary format is 2. To perform multiplication of two floating point numbers $\pm s1 \times b^{e1}$ and $\pm s2 \times b^{e2}$, the significant or mantissa parts are multiplied to get the product mantissa and exponents are added to get the product exponent. i.e.; the product is $\pm(s1 \times s2) \times b^{(e1+e2)}$. The hardware block diagram of floating point multiplier is shown in fig. 2.

The important blocks in the implementation of proposed floating point multiplier [3] is described below.

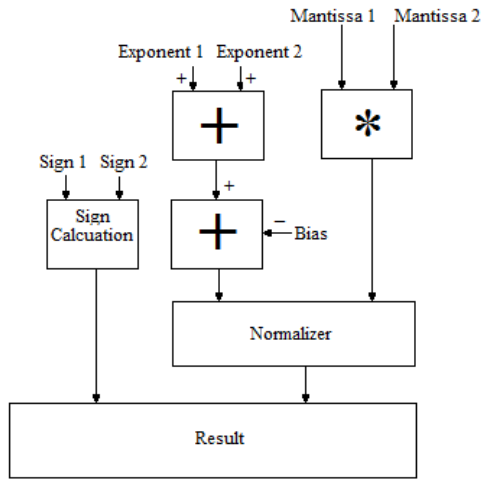

Fig. 2 Floating point multiplier

## A. Sign Calculation

The MSB of floating point number represents the sign bit. The sign of the product will be positive if both the numbers are of same sign and will be negative if numbers are of opposite sign. So, to obtain the sign of the product, we can use a simple XOR gate as the sign calculator.

## B. Addition of Exponents

To get the product exponent, the input exponents are added together. Since we use a bias in the floating point format exponent, we need to subtract the bias from the sum of exponents to get the actual exponent. The value of bias is $127_{10}$ ($01111111_2$) for single precision format and $1023_{10}$($01111111111_2$) for double precision format. In proposed custom precision format also, a bias of 127 is used.

The computational time of mantissa multiplication operation is much more that the exponent addition. So a simple ripple carry adder and ripple borrow subtracter is optimal for exponent addition.

## C. Karatsuba-Urdhva Tiryagbhyam binary multiplier

In floating point multiplication, most important and complex part is the mantissa multiplication. Multiplication operation requires more time compared to addition. And as the number of bits increase, it consumes more area and time. In double precision format, we need a 53x53 bit multiplier and in single precision format we need 24x24 bit multiplier. It requires much time to perform these operations and it is the major contributor to the delay of the floating point multiplier. To make the multiplication operation more area efficient and faster, the proposed model uses a combination of Karatsuba algorithm and Urdhva Tiryagbhyam algorithm.

Karatsuba algorithm uses a divide and conquer approach where it breaks down the inputs into Most Significant half and Least Significant half and this process continues until the operands are of 8-bits wide. Karatsuba algorithm is best suited for operands of higher bit length. But at lower bit lengths, it is not as efficient as it is at higher bit lengths. To eliminate this problem, Urdhva Tiryagbhyam algorithm is used at the lower stages. The model of Urdhva-Tiryagbhyam algorithm is shown in Fig. 3.

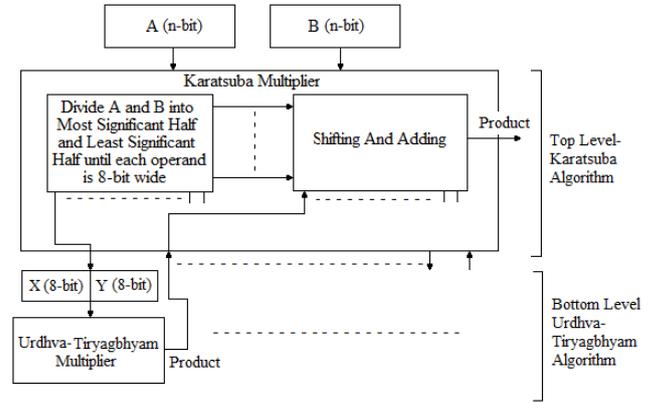

Fig. 3 Karatsuba-Urdhva multiplier model

Urdhva Tiryagbhyam algorithm is the best algorithm for binary multiplication in terms of area and delay. But as the number of bits increases, delay also increases as the partial products are added in a ripple manner. For example, for 4-bit multiplication, it requires 6 adders connected in a ripple manner. And 8-bit multiplication requires 14 adders and so on. Compensating the delay will cause increase in area. So Urdhva Tiryagbhyam algorithm is not that optimal if the number of bits is much more. If we use Karatsuba algorithm at higher stages and Urdhva Tiryagbhyam algorithm at lower stages, it can somewhat compensate the limitations in both the algorithms and hence the multiplier becomes more efficient. The circuit is further optimized by using carry select and carry save adders instead of ripple carry adders. This reduces the delay to a great extent with minimal increase in hardware. These two algorithms are explained in detail in the below sections.

*Urdhva Tiryagbhyam algorithm for multiplication*

Urdhva-Tiryagbhyam sutra is an ancient Vedic mathematics method for multiplication [4, 5, 6, 7]. It is a general formula applicable to all cases of multiplication. The formula is very short and consists of only one compound word and means 'Vertically and crosswise'. In Urdhva Tiryagbhyam algorithm, the number of steps required for multiplication can be reduced and hence the speed of multiplication is increased.

An illustration of steps for computing the product of two 4-bit numbers is shown below [8, 9]. The two input are $a_3a_2a_1a_0$ and $b_3b_2b_1b_0$ and let the $p_7p_6p_5p_4p_3p_2p_1p_0$ be the product. And the temporary partial products are $t_0, t_1, t_2, \ldots, t_6$.

The partial products are obtained from the steps given below. The line notation of the steps is shown in Fig. 4.

Step1: $t_0(1 bit) = a_0 b_0$.
Step2: $t_1(2 bit) = a_1 b_0 + a_0 b_1$.
Step3: $t_2(2 bit) = a_2 b_0 + a_1 b_1 + a_0 b_2$
Step4: $t_3(3 bit) = a_3 b_0 + a_2 b_1 + a_1 b_2 + a_0 b_3$.
Step5: $t_4(2 bit) = a_3 b_1 + a_2 b_2 + a_1 b_3$.
Step6: $t_5(2 bit) = a_3 b_2 + a_2 b_3$.
Step7: $t_6(1 bit) = a_3 b_3$

The product is obtained by adding $s_1, s_2$ and $s_3$ as shown below, where $s_1, s_2$ and $s_3$ are the partial sum obtained.

$s_1 = t_6\ t_5[0]\ t_4[0]\ t_3[0]\ t_2[0]\ t_1[0]\ t_0$
$s_2 = t_5[1]\ t_4[1]\ t_3[1]\ t_2[1]\ t_1[1]$
$s_3 = t_3[2]$

$$\begin{array}{r} \text{Product} = t_6\ t_5[0]\ t_4[0]\ t_3[0]\ t_2[0]\ t_1[0]\ t_0\ + \\ t_5[1]\ t_4[1]\ t_3[1]\ t_2[1]\ t_1[1]\ 0\ + \\ t_3[2]\quad 0\quad 0\quad\ \ 0\quad\ \ 0 \\ \hline p_7\ p_6\ p_5\quad p_4\quad p_3\quad p_2\quad p_1\quad p_0 \end{array}$$

The expressions for product bits are as shown below.
$p_0 = a_0 b_0$
$p_1 = LSB\ of\ (Sum(ADDER\ 1))$
$\quad = LSB\ of\ (a_1 b_0 + a_0 b_1)$
$p_2 = LSB\ of\ (Sum(ADDER\ 2))$
$\quad = LSB\ of\ (MSB(ADDER1) + a_2 b_0 + a_1 b_1 + a_0 b_2)$
$p_3 = LSB\ of\ (Sum(ADDER\ 3))$
$\quad = LSB\ of\ (MSB(ADDER\ 2) + a_3 b_0 + a_2 b_1 + a_1 b_2 + a_0 b_3)$
$p_4 = LSB\ of\ (Sum(ADDER\ 4))$
$\quad = LSB\ of\ (MSB(ADDER1) + a_3 b_1 + a_2 b_2 + a_1 b_3)$
$p_5 = LSB\ of\ (Sum(ADDER\ 5))$
$\quad = LSB\ of\ (MSB(ADDER1) + a_3 b_2 + a_2 b_3)$
$p_6 = LSB\ of\ (Sum(ADDER\ 6))$
$\quad = LSB\ of\ (MSB(ADDER1) + a_3 b_3)$
$p_7 = Carry\ of\ ADDER$

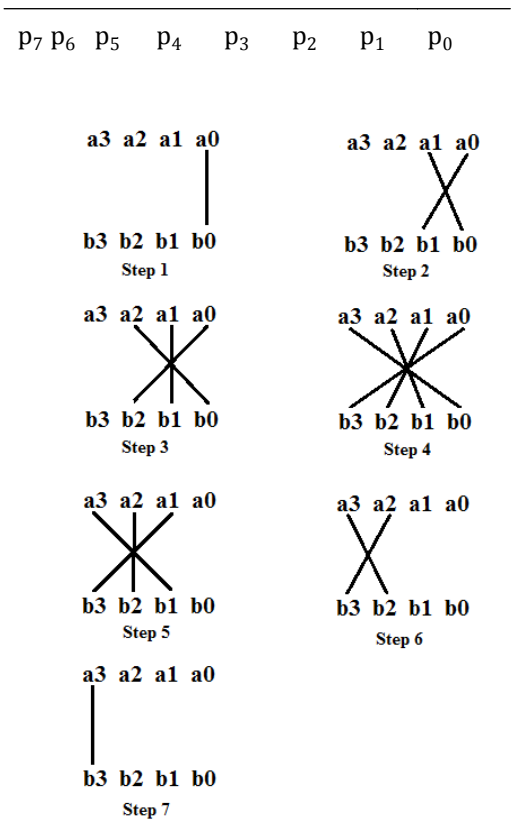

Fig. 4 Line notation of Urdhva Tiryagbhyam sutra

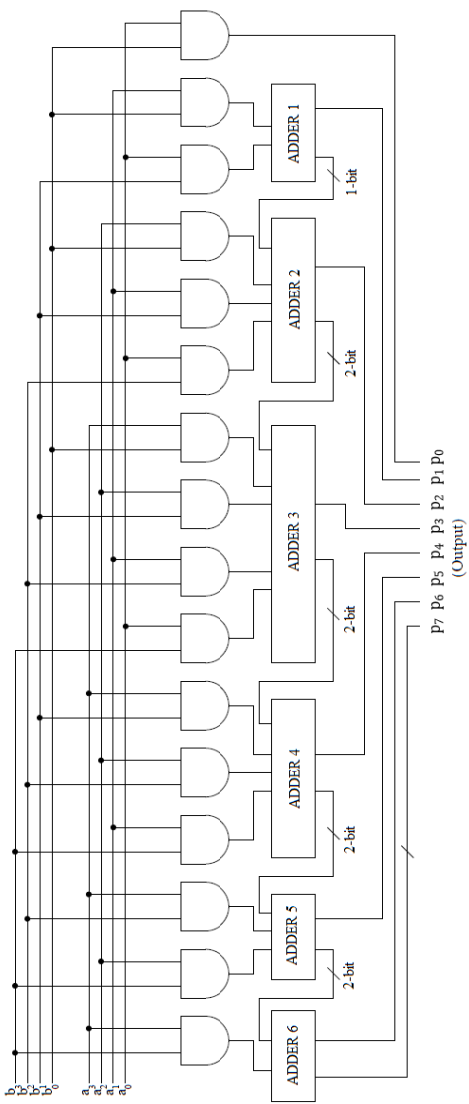

Fig. 5 Hardware architecture for 4x4 Urdhva Tiryagbhyam multiplier.

This method can be further optimized to reduce the number of hardware. A more optimized hardware architecture [9, 10] is shown in Fig. 5. This model actually helps to eliminate the need for three operand 7-bit adder and hence reduces hardware and delay. The adders are connected in ripple manner.

Since there are more than two operands in adders 2 to 5, we can use carry save addition to implement adders 2 to 5. This technique reduces the delay to a great extend compared to the ripple carry adder.

*Karatsuba Algorithm for multiplication*

Karatsuba multiplication algorithm [11, 12] is best suited for multiplying very large numbers. This method is discovered by Anatoli Karatsuba in 1962. It is a divide and conquer method, in which we divide the numbers into their Most Significant half and Least Significant half and then multiplication is performed.

Karatsuba algorithm reduces the number of multipliers required by replacing multiplication operations by addition operations. Additions operations are faster than multiplications and hence the speed of multiplier is increased. As the number of bits of inputs increase, Karatsuba algorithm becomes more efficient. This algorithm is optimal if width of inputs is more than 16 bits. The hardware architecture of Karatsuba algorithm is shown in fig. 6. Karatsuba algorithm for two inputs X and Y can be explained as follow.

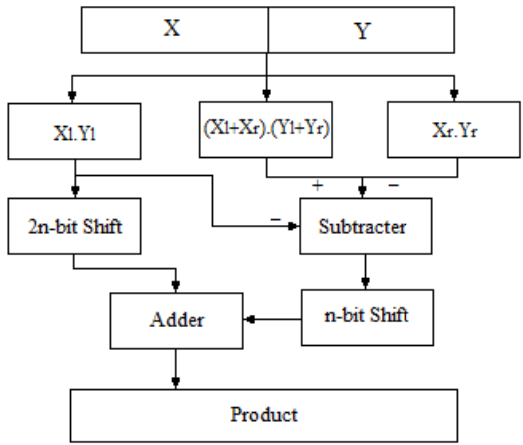

Fig. 6 Karatsuba multiplier

Product= $X.Y$
X and Y can be written as,
$$X = 2^{n/2}. X_l + X_r \quad (1)$$
$$Y = 2^{n/2}. Y_l + Y_r \quad (2)$$
Where $X_l, Y_l$ and $X_r, Y_r$ are the Most Significant half and Least Significant half of X and Y respectively, and n is the number of bits.
Then,
$$X.Y = \left(2^{\frac{n}{2}}. X_l + X_r\right).(2^{\frac{n}{2}}. Y_l + Y_r)$$
$$= 2^n. X_l Y_l + 2^{n/2} (X_l Y_r + X_r Y_l) + X_r Y_r \quad (3)$$

The Second term in equation (3) can be optimized to reduce the number of multiplication operations.

i.e.; $X_l Y_r + X_r Y_l = (X_l + X_r)(Y_l + Y_r) - X_l Y_l - X_r Y_r$ (4)

The equation (3) can be re-written as,
$$X.Y = 2^n. X_l Y_l + X_r Y_r + 2^{\frac{n}{2}} ((X_l + X_r)(Y_l + Y_r) - X_l Y_l - X_r Y_r) \quad (5)$$

The recurrence of Karatsuba algorithm is,
$$T(n) = 3T\left(\frac{n}{2}\right) + O(n) \quad O(n^{1.585})$$

### D. Normalization of the result

Floating point representations have a hidden bit in the mantissa, which always has a value 1 and hence it is not stored in the memory to save one bit. A leading 1 in the mantissa is considered to be the hidden bit, i.e. the 1 just immediate to the left of decimal point. Usually normalization is done by shifting, so that the MSB of mantissa becomes nonzero and in radix 2, nonzero means 1. The decimal point in the mantissa multiplication result is shifted left if the leading 1 is not at the immediate left of decimal point. And for each left shift operation of the result, the exponent value is incremented by one. This is called normalization of the result. Since the value of hidden bit is always 1, it is called 'hidden 1'.

### E. Representation of exceptions

Some of the numbers cannot be represented with a normalized significand. To represent those numbers a special code is assigned to it. In the proposed model, we use four output signals namely Zero, Infinity, NaN (Not-a-number) and Denormal to represent these exceptions. If the product has $exponent + bias = 0$ and $significand = 0$, then the result is taken as Zero (±0). If the product has $exponent + bias = 255$ and $significand = 0$, then the result is taken as Infinity (∞). If the product has $exponent + bias = 255$ and $significand \neq 0$, then the result is taken as NaN. Denormalized values or Denormals are numbers without a hidden 1 and with the smallest possible exponent. Denormals are used to represent certain small numbers that cannot be represented as normalized numbers. If the product has $exponent + bias = 0$ and $significand \neq 0$, then the result is represented as Denormal. Denaormal is represented as $\pm 0.s \times 2^{-126}$, where s is the significand.

### III. IMPLIMENTATION AND RESULTS

The main objective of this paper is to design and implement a floating point multiplier which must be efficient in its operation both in terms of delay and area. Since mantissa multiplication is the most complex part in the floating point multiplier, we designed a multiplier which can operate at high speed and increase in delay and area is significantly less with increase in number of bits. Floating point multiplier with IEEE-754 standard format is implemented using Verilog HDL and tested. The multiplier units are further optimized by replacing simple adders with efficient adders like carry select adders and carry save adders. The model is synthesized and simulated using Xilinx Synthesis Tools (ISE 14.7) targeted on Saprtan-3E and Virtex-4 fpga. The summary of results on Virtex-4 fpga is given in table I and table II. Comparison with various multiplier units is given in tables III, IV, V, VI and VII.

TABLE I
Performance analysis of Karatsuba-Urdhva multipliers

|  | 8-bit multiplier | 16-bit multiplier | 24-bit multiplier | 32-bit multiplier |
|---|---|---|---|---|
| Slices | 113 | 410 | 972 | 1389 |
| LUTs | 120 | 451 | 1018 | 1545 |
| IOBs | 33 | 65 | 97 | 129 |
| Delay | 9.396ns | 11.514ns | 12.996ns | 13.141ns |
| $f_{max}$ (MHz) | 274.469 | 248.964 | 226.508 | 209.606 |
| Logic levels | 14 | 22 | 31 | 39 |

TABLE II
Performance analysis of Floating point multipliers in the proposed model.

|  | Slices | LUTs | IOBs | Delay (ns) | $f_{max}$ (MHz) | Max. comb. path delay(ns) |
|---|---|---|---|---|---|---|
| Single precision | 977 | 1073 | 97 | 16.182 | 226.508 | 9.831 |
| Double precision | 3877 | 4033 | 193 | 18.966 | 173.952 | 10.736 |

TABLE III
Delay comparison of various 8-bit multipliers with proposed Karatsuba-Urdhva multiplier

|  | Ref. [8] | Ref. [9] | Ref. [13] | Proposed multiplier |
|---|---|---|---|---|
| Width | 8-bit | 8-bit | 8-bit | 8-bit |
| Delay | 28.27ns | 15.050ns | 23.973ns | 9.396ns |

TABLE IV
Delay comparison of various 16-bit multipliers with proposed Karatsuba-Urdhva multiplier

|  | Ref. [14]-vedic multiplier | Ref. [7] | Proposed multiplier |
|---|---|---|---|
| Width | 16-bit | 16-bit | 16-bit |
| Delay | 13.452ns | 27.148ns | 11.514ns |

TABLE V
Delay and area comparison of 24-bit multipliers with proposed Karatsuba-Urdhva multiplier

|  | Slices | LUTs | Delay |
|---|---|---|---|
| Ref. [15] | 1306 | 2329 | 16.316ns |
| Proposed multiplier | 972 | 1018 | 12.996ns |

TABLE VI
Delay and area comparison of 32-bit multipliers with proposed Karatsuba-Urdhva multiplier

|  | LUTs | Delay |
|---|---|---|
| Ref. [14]- Modified Booth multiplier (Radix-8) | 2721 | 12.081ns |
| Ref. [14]- Modified Booth multiplier (Radix-16) | 7161 | 11.564ns |
| Ref. [14] | 2704 | 9.536ns |
| Proposed multiplier | 1545 | 13.141ns |

TABLE VII
Delay and area comparison of SP-floating point multiplier with proposed SP FP multiplier

|  | Slices | LUTs | Delay |
|---|---|---|---|
| Ref. [15] | 1269 | 2270 | 18.783ns |
| Ref. [3] | 1149 | 1146 | -- |
| Proposed multiplier | 977 | 1073 | 16.182ns |

IV. CONCLUSION AND FUTURE WORK

This paper shows how to effectively reduce the percentage increase in delay and area of a floating point multiplier by using a very efficient combination of Karatsuba and Urdhva-Tiryagbhyam algorithms. The model can be further optimized in terms of delay by using pipelining methods and precision of the result can be increased by adding efficient truncation and rounding methods.